\documentclass[twocolumn]{aastex631}

\usepackage{amsmath}

\begin{document}

\title{Half a Million Binary Stars identified from the low resolution spectra of LAMOST}

\correspondingauthor{Yingjie Jing}
\email{jyj@nao.cas.cn}
\correspondingauthor{Jie Wang}
\email{jie.wang@nao.ac.cn}

\author{Yingjie Jing}
\affiliation{National Astronomical Observatories, Chinese Academy of Sciences, Beijing, 100101, China}

\author{Tian-Xiang Mao}
\affiliation{National Astronomical Observatories, Chinese Academy of Sciences, Beijing, 100101, China}

\author{Jie Wang}
\affiliation{National Astronomical Observatories, Chinese Academy of Sciences, Beijing, 100101, China}
\affiliation{Institute for Frontiers in Astronomy and Astrophysics, Beijing Normal University, Beijing 102206, China}
\affiliation{University of Chinese Academy of Sciences, 19 A Yuquan Rd, Shijingshan District, Beijing, 100049, China}

\author{Chao Liu}
\affiliation{Key Laboratory of Space Astronomy and Technology, National Astronomical Observatories, Chinese Academy of Sciences, Beijing 100101, China}
\affiliation{University of Chinese Academy of Sciences, 19 A Yuquan Rd, Shijingshan District, Beijing, 100049, China}
\affiliation{Institute for Frontiers in Astronomy and Astrophysics, Beijing Normal University, Beijing 102206, China}

\author{Xiaodian Chen}
\affiliation{CAS Key Laboratory of Optical Astronomy, National Astronomical Observatories, Chinese Academy of Sciences, Beijing 100101, China}
\affiliation{University of Chinese Academy of Sciences, 19 A Yuquan Rd, Shijingshan District, Beijing, 100049, China}
\affiliation{Institute for Frontiers in Astronomy and Astrophysics, Beijing Normal University, Beijing 102206, China}



\begin{abstract}

Binary stars are prevalent yet challenging to detect. We present a novel approach using convolutional neural networks (CNNs) to identify binary stars from low-resolution spectra obtained by the LAMOST survey. The CNN is trained on a dataset that distinguishes binaries from single main sequence stars based on their positions on the Hertzsprung-Russell diagram. Specifically, the training data labels stars with mass ratios between approximately 0.71 and 0.93 as intermediate mass ratio binaries, while excluding those beyond this range. The network achieves high accuracy with an area under the receiver operating characteristic curve of 0.949 on the test set. Its performance is further validated against known eclipsing binaries (97\% detection rate) and binary stars identified by radial velocity variations (92\% detection rate). Applying the trained CNN to a sample of one million main sequence stars from LAMOST DR10 and Gaia DR3 yields a catalog of 468,634 binary stars, which are mainly intermediate mass ratio binaries given the training data. This catalog includes 115 binary stars located beyond 10 kpc from the Sun and 128 cross-matched with known exoplanet hosts from the NASA Exoplanet Archive. This new catalog provides a valuable resource for future research on the properties, formation, and evolution of binary systems, particularly for statistically characterizing large populations.

\end{abstract}




\section{Introduction} \label{sec:intro}

Binary and multiple star systems are ubiquitous, comprising roughly half of all stars.  Understanding these systems is fundamental to various areas of astronomy.  For instance, precise measurements of mass, radius, and luminosity are possible for stars in binary systems \citep[e.g.][]{Torres_2010, Eker_2018}.  Furthermore, binaries are crucial for testing theories of stellar formation and evolution, galactic archaeology, gravitational waves, and high-energy astrophysics \citep[e.g.][]{Duquennoy_1991, Raghavan_2010, Duch_ne_2013,Moe_2017,Breivik_2019,Tauris_2023}.  Some binary systems even serve as valuable distance indicators \citep{Kang2007, Pietrzyski2013}. Therefore, the availability of a larger sample of binary systems is indispensable and advantageous for conducting comprehensive studies in the aforementioned fields.

Traditional methods for identifying binary stars include detecting variations in radial velocity \citep[e.g.][]{Pryor_1988,Cote_1994,Cote_1996} or brightness (light curves) \citep[e.g.][]{Yan_1994,Albrow_2001,Milone_2012}. The Kepler mission, using light curve data, has been particularly successful in identifying thousands of eclipsing binaries \citep[e.g.,][]{Kirk2016, Abdul-Masih2016}. However, these methods often require multiple observations over time and are most effective for bright, close binaries with short orbital periods.  Analyzing the color-magnitude diagram (CMD) provides another approach, as binaries tend to be brighter and redder than single main-sequence stars of the same mass \citep[e.g.][]{Sollima_2010,Milone_2012,Li_2013}.  Sophisticated statistical techniques applied to CMDs have improved the precision of this method \citep[e.g.][]{Hurley_1998,Naylor_2006,Sarro_2014,Sheikhi_2016,Li_2017,Liu2019,Li_2020}. Despite these advancements, the number of identified binaries remains limited.

Large spectroscopic surveys such as APOGEE \citep{Holtzman_2015}, RAVE \citep{Steinmetz_2006}, LAMOST \citep{Cui2012,Zhao2012} offer new opportunities for binary star research. These surveys enable the identification of spectroscopic binaries (SBs) through Doppler shifts in their spectral lines \citep[see review by][]{Merle_2020}.  The composite spectra of double-lined spectroscopic binaries (SB2s) provide additional clues. SBs cover a wide range of stellar masses and orbital periods \citep{Soderhjelm_2004}, and their detection is less affected by distance, making them well-suited for large-scale studies. Previous work using these spectroscopic surveys has resulted in binary star catalogs ranging in size from hundreds \citep[RAVE, e.g.][]{Matijevic2010, Birko019AJ} to thousands \citep[APOGEE, e.g.][]{Price-Whelan2018,El-Badry2018,Kounkel2021AJ}. Studies using low-resolution LAMOST spectra have identified hundreds of thousands of binaries \citep[e.g.,][]{Liu20204,Qian2019,Liu_Hao-Bin2024}. \citet{Zhangbo2022} find several thousand binary stars from LAMOST median resolution spectra. Similarly, Gaia data \citep{Gilmore_2012,Gaia2016, Gaia2023} have led to even larger catalogs, with over a million binaries identified using Gaia eDR3 \citep{El-Badry2021}.

Spectroscopic binaries exhibit distinctive spectral features that enable direct identification through analysis of their stellar spectra. While conventional methods often rely on fitting observed spectra to template spectra \citep[e.g.,][]{El-Badry2018,Traven2020}, neural networks offer a more powerful and efficient alternative \citep[see reviews by][]{Lecun2015,Goodfellow2016}.  Neural networks excel at complex tasks by learning intricate patterns from data, eliminating the need for manual feature engineering. This data-driven approach is particularly well-suited for binary star classification. In this work, we leverage the large volume of low-resolution stellar spectra from the LAMOST survey \citep{Cui2012,Zhao2012} and employ convolutional neural networks (CNNs) \citep[e.g.,][]{CNN,CNN2} to identify binary stars. CNNs and other artificial neural networks have been successfully applied to various astronomical problems \citep[e.g.,][]{Ting2018, Davies2019, Pearson2019}, and we expect them to be effective in extracting relevant information from spectral features for binary star detection.

The paper is organized as follows: Section \ref{sect:data} details the dataset along with the network model. Section \ref{sect:res} reviews our model's performance and offers a catalog of main sequence binary stars. Section \ref{sect:con} summarizes the outcomes of our study.

\section{Data and Methods}
\label{sect:data}

\subsection{Dataset}
\begin{figure}
	\includegraphics[width=\columnwidth]{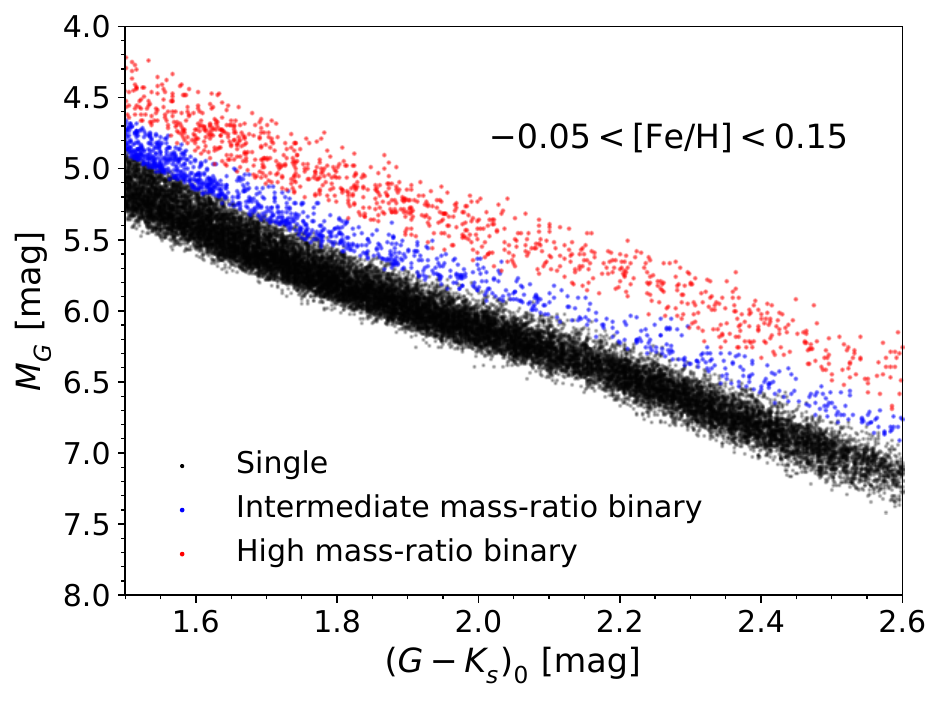}
	\caption{Color-magnitude diagram for stars with $-0.05< \mathrm{[Fe/H]} <0.15$ from the train sample. The black, blue, and red points represent identified single main sequence, intermediate, and high mass-ratio binary stars, respectively. The single stars and intermediate mass ratio binaries are selected as training sample. The G-band absolute magnitude ($M_G$) is plotted against the dereddened color index $(G-K_s)_0$. }
	\label{fig:HR_train}
\end{figure}

In order to train the neural network, a larger number of binary and single star spectra are needed. We utilized the star sample from the \citet{Liu2019} which selected a solar neighborhood stars sample and identified the single and binary stars based on the fact that binary sequence is located above the single main sequence in Hertzprung-Russell (HR) diagram. For illustrative purposes, we present a color-magnitude diagram (CMD) for stars within a [Fe/H] range ($-0.05 < \text{[Fe/H]} < 0.15$) from this sample in Figure~\ref{fig:HR_train}. Here, CMD is the dereddened color index $(G - K_s)_0$ compared to the absolute magnitude of the G band $M_G$. Unresolved binary stars exhibit a brighter absolute magnitude compared to single stars. With a larger shift $\Delta M_G = M_G - M_{G,\mathrm{single}}$, the probability that the star is a binary increases, as does the likelihood of a high mass ratio if it is binary (see also Figure 9 in \citet{Liu2019}).

Specifically, we labeled stars with $\Delta M_G$ values between $-0.25$ and $0.25$ as single stars (represented by black points in Figure~\ref{fig:HR_train}).  Stars with $\Delta M_G$ values between $-0.5$ and $-0.25$ (shown in blue points) were labeled as binaries (corresponding to an intermediate mass ratios between approximately 0.71 and 0.93). The threshold of $\Delta M_G = -0.25$ was chosen to ensure the purity of the binary sample, accounting for magnitude uncertainties. High-mass-ratio binaries ($\Delta M_G < -0.5$, red points) were excluded from our training set because their spectra are very similar to single star spectra, making spectral-based discrimination challenging. Following these selection criteria, our final training sample consisted of 68,299 single stars and 3,818 binary stars.  These stars were subsequently divided into training, validation, and test sets, as described in the following subsection.

We constructed our initial stellar sample by cross-matching stars  from LAMOST DR10 low-resolution A, F, G and K stars catalog (\url{https://www.lamost.org/dr10/}), Gaia DR3
\citep{Gaia2016, Gaia2023}, and 2MASS \citep{Skrutskie2006} following the procedure outlined in \citet{Liu2019}. This initial sample includes about 7 million spectra. The LAMOST low-resolution spectra span a wavelength range of 3700 to 9000 \AA{}, with a resolution of $\mathrm{R} \approx 1800 $ \citep{Cui2012, Luo2012}.

The train sample from \citet{Liu2019} satisfies the following selection criteria:

\begin{enumerate}
    \item $3800< T_{\mathrm{eff}} < 6500$ K;
    \item log $g > 4$;
    \item $1.5 < (G-K_s)_0 < 2.6$;
    \item signal-to-noise ratio at $g$ band of LAMOST spectra is larger than
$20$;
    \item $\varpi >3$ so that most of the stars are located within $\sim 300 ~\mathrm{pc}$ (excluded in this work); 
    \item G-band absolute magnitude $M_\textrm{G} > 4$mag (excluded in this work).
\end{enumerate}

As our network relies exclusively on spectra for input, we have excluded selection criteria 5 and 6. Consequently, our final main sequence stars sample comprises 1,258,912 spectra associated with 971,805 stars, many of which have been observed multiple times.

\subsection{Spectra Data Preprocessing}

Before generating the training set, various data preprocessing steps are taken to enhance the network's performance. Initially, data cleaning is applied to eliminate spectra with excessively masked data points. Subsequently, each flux spectrum is interpolated onto a new wavelength grid and normalized by dividing by the corresponding smoothed flux $f_s(\lambda)$ \citep{Ho2017}. The definition of smoothed flux $f_s(\lambda)$ is

\begin{equation}
\label{eq:smooth}
f_s\left(\lambda\right)=\frac{\sum_{i}\left(f_{i} w_{i}\left(\lambda\right)\right)}{\sum_{i}\left(w_{i}\left(\lambda\right)\right)},
\end{equation}

where $f_i$ is the flux at $\lambda_i$ and the weight $w_i(\lambda)$ is a
Gaussian function

\begin{equation}
\label{eq:gau}
w_{i}\left(\lambda\right)=e^{-\frac{\left(\lambda-\lambda_{i}\right)^{2}}{\sigma^{2}}}.
\end{equation}
The $\sigma$ is set to $35~\mathrm{nm}$ to eliminate large-scale variations. We then use the entire spectrum, with wavelengths ranging from 3700 to 9000 \AA{}, as input to the network. 

We randomly divide the aforementioned data set in an 80\%, 10\%, 10\% split to form the training, validation, and test sets. In particular, since the proportion of binary stars to single stars is approximately $1:18$ in our training set, this classification poses an imbalance issue. The impact of class imbalance on classification performance is both harmful and intricate \citep{Buda2017}. To mitigate this impact, we adopt the method described in \citet{Buda2017} to oversample binary stars to match the number of single stars by repeatedly sampling within the training set.

\subsection{Deep learning model}

The basic structure of CNNs is named the convolutional layer, which
is defined as
\begin{equation}
\label{eq:layer}
\boldsymbol{x}_{n}^{l}=a\left(\sum_{k} \mathbf{W}_{n}^{l} \otimes \boldsymbol{x}_{n-1}^{k}+\boldsymbol{b}_{n}^{l}\right).
\end{equation}

Here, $\otimes$ indicates the convolution operation, $\mathbf{W}_{n}^{l}$
denotes the trainable convolutional kernels of layer $n$, $l$ indicates the $l$-th
kernel in this layer and $k$ indicates the output corresponding to the $k$-th
convolutional kernel of the previous layer. Since we hope to classify the binary stars from the spectra data, the convolution operation
$\otimes$ means 1-D convolution in this work.

The architecture of our network includes a basic residual learning block (refer to Figure 2 in \citet{He2015}) followed by two fully connected layers. To mitigate the risk of overfitting due to the limited number of binary star samples, we integrate dropout layers \citep{Srivastava2014} within the fully connected layers at a rate of 0.5 and employ early stopping during training. We utilize the cross-entropy loss function and set the learning rate as $1e^{-4}$. For each spectrum, the trained network estimates the probability $p_{\rm b}$ of it being a binary star, with the value of $p_{\rm b}$ lying between 0 and 1.

\section{Results}
\label{sect:res}
In this section, we will assess the network's performance using the test set and known binary stars samples from previous studies. These binary star samples include eclipsing binaries and binary stars identified from variations in radial velocity. We then apply the network to the entire sample.

\subsection{Test set}
\begin{figure}
	\includegraphics[width=\columnwidth]{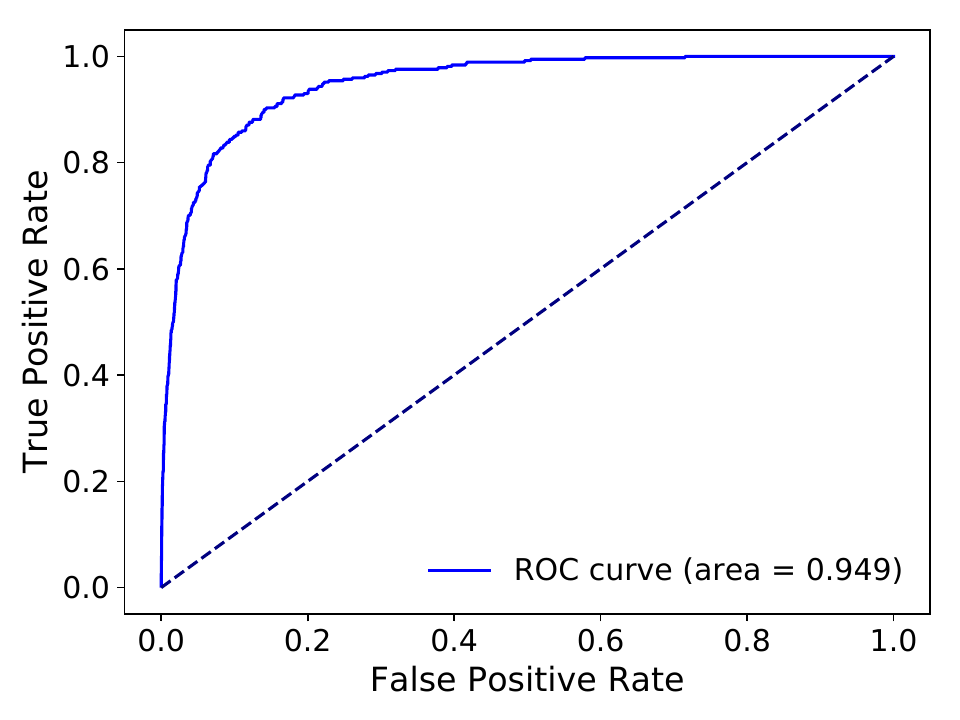}
	\caption{Receiver Operating Characteristic (ROC) curve for the test set. The solid blue line represents the performance of the trained CNN model, while the dashed orange line indicates the ROC curve for random selection. The area under the curve (AUC) for the model is 0.949, demonstrating significantly better performance than random classification.}
	\label{fig:roc}
\end{figure}

The receiver operating characteristic (ROC) curve is frequently utilized to assess the efficacy of a binary classification system. A larger area under the ROC curve indicates superior performance by the classifier. The ROC curve is a graph plotting the true positive rate (TPR) versus the false positive rate (FPR) at various thresholds. In Figure~\ref{fig:roc}, the ROC curve for the test set is represented by a solid line. For comparison, a dashed line denotes the ROC curve for a random sample, which has an area of 0.5. The ROC curve of the test set is significantly better than that of random selection, covering a larger area (0.949).

\begin{figure}
	\includegraphics[width=\columnwidth]{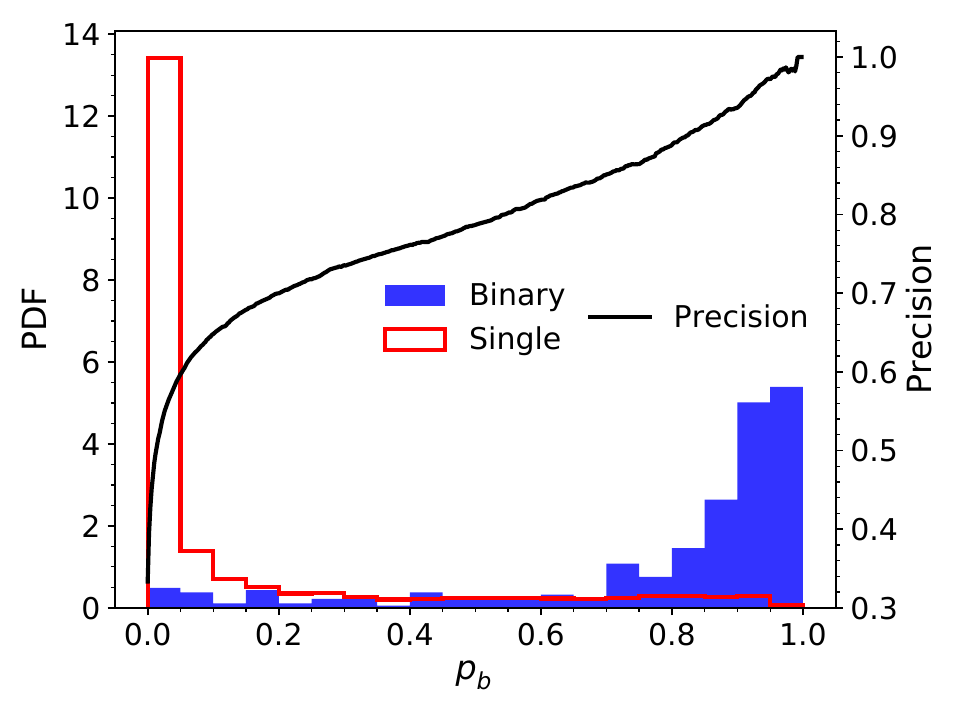}
	\caption{Probability density function (PDF) of the predicted probabilities ($p_b$) for binary (blue region) and single (red line) stars in the test set. The black line shows the precision as a function of the adopted probability cut-off threshold, assuming a binary-to-single star ratio of 1:2.
    }
	\label{fig:test_prob}
\end{figure}

We present the probability density distribution function (PDF) of $p_{\rm b}$ for both single and binary stars in the test set in Figure~\ref{fig:test_prob}. The majority of single stars (95.5\%) exhibit a $p_b$ less than 0.2, while most binary stars (73\%) demonstrate a $p_b$ greater than 0.8. This indicates that the network performs well in classifying star spectra. To ensure the binary stars identified by the network are as accurate as possible, we estimate the precision, which is dependent on the threshold cut-off of $p_b$, denoted as $p_\mathrm{th}$, and the binary-to-single star ratio $r$ in the input sample. The precision is defined as
\begin{equation}
    \mathrm{Precision} = \frac{N_{\rm TP}}{N_{\rm TP} + N_{\rm FP}\times N_{\rm b}/N_{\rm s}/r},
\end{equation}
where $N_{\rm b}$ and $N_{\rm s}$ are the number of binary and single stars in our test set, respectively, and $N_{\rm TP}$ and $N_{\rm FP}$ represent the number of true positive (correctly identified binary) and false positive (single star misclassified as binary) classifications with $p_{\rm b} > p_\mathrm{th}$, respectively. By fixing the binary-to-single ratio at $1:2$, we illustrate the precision as a function of $p_\mathrm{th}$ in the black curve of Figure~\ref{fig:test_prob}. Our results illustrate that the precision reaches 0.79 (0.89) when $p_\mathrm{th}$ is set to 0.5 (0.8).

\begin{figure}
	\includegraphics[width=\columnwidth]{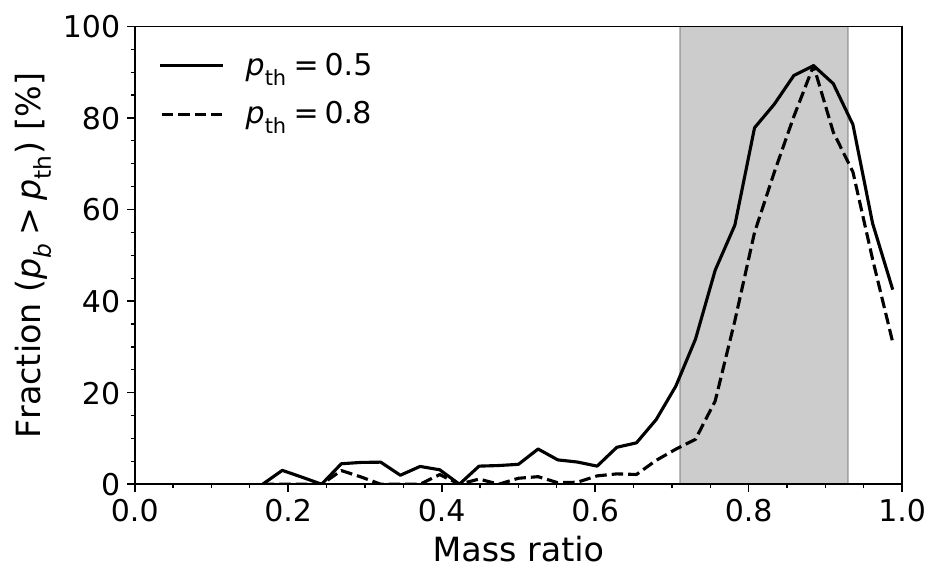}
	\caption{The fraction of binary stars as a function of mass ratio for different $p_\mathrm{th}$ values. The shaded area denotes the regions of intermediate-mass-ratio binaries (corresponding to mass ratios between approximately 0.71 and 0.93) which is included in the train sample.}
	\label{fig:prob_mass-ratio}
\end{figure}

It is also worth to investigate how the network's sensitivity changes with different mass ratios. Figure~\ref{fig:prob_mass-ratio} displays the fraction of stars classified as binaries as a function of their mass ratio driven by \citet{Liu2019}, for different probability thresholds ($p_\mathrm{th}$). The stars shown in this figure are from the test set and high-mass-ratio stars from \citet{Liu2019}. A trend influenced by the training data is evident. The binary fraction is highest for mass ratios around 0.71-0.93, which is indicated by the shaded region and corresponds to the intermediate-mass-ratio binaries used in the training process. The figure shows a decrease in the binary fraction as mass ratios close to 1. This is partially a result of the exclusion of high-mass-ratio binaries ($\Delta M_G < -0.5$) from the training data, which is due to the intrinsic similarity of their spectra to those of single stars. For low mass ratios, the sensitivity decreases significantly. This is likely due to the training data using specific cuts on $\Delta M_G$ to distinguish between single stars and intermediate-mass-ratio binaries. While we expect that a more comprehensive training sample might improve the detection rate for lower mass ratio, the detection of highest mass ratio binaries may remain difficult due to the limitations in spectrally distinguishing them from single stars. The overall fraction of classified binaries is also shown to be lower for curves with higher $p_\mathrm{th}$ values across all mass ratios.

\subsection{Comparison with other methods}

\begin{figure}
	\includegraphics[width=\columnwidth]{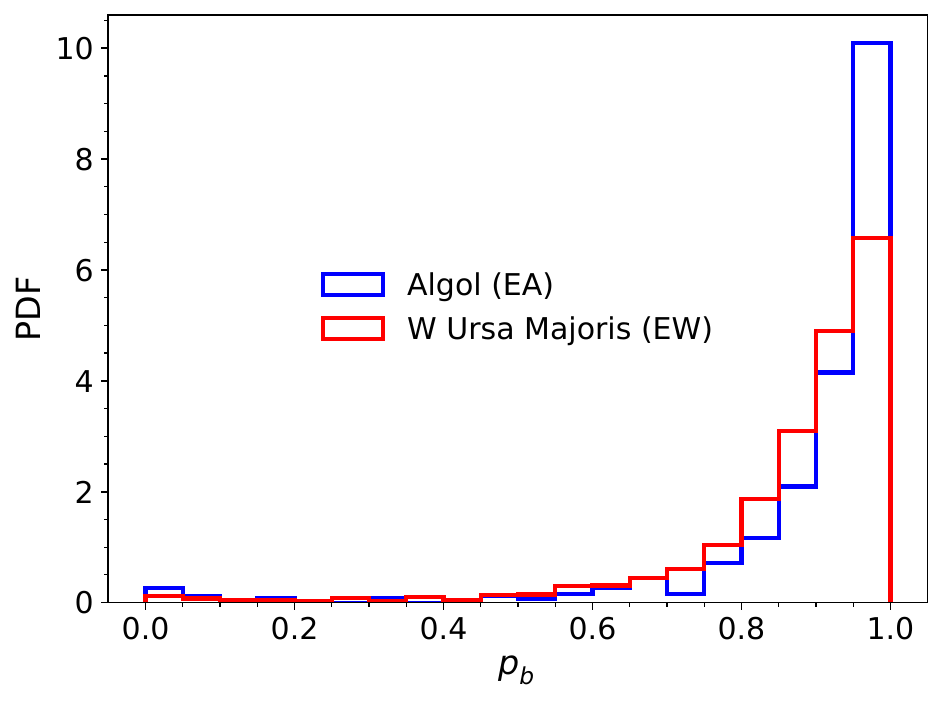}
	\caption{Probability density function (PDF) of the predicted binary probabilities ($p_b$) for eclipsing binary stars. The blue solid line corresponds to Algol-type (EA) binaries, while the red solid line corresponds to W Ursa Majoris-type (EW) binaries, demonstrating a high detection rate of 96.45\% for EA and 96.40\% for EW binaries.}
	\label{fig:EWL_EAL}
\end{figure}

\begin{figure}
	\includegraphics[width=\columnwidth]{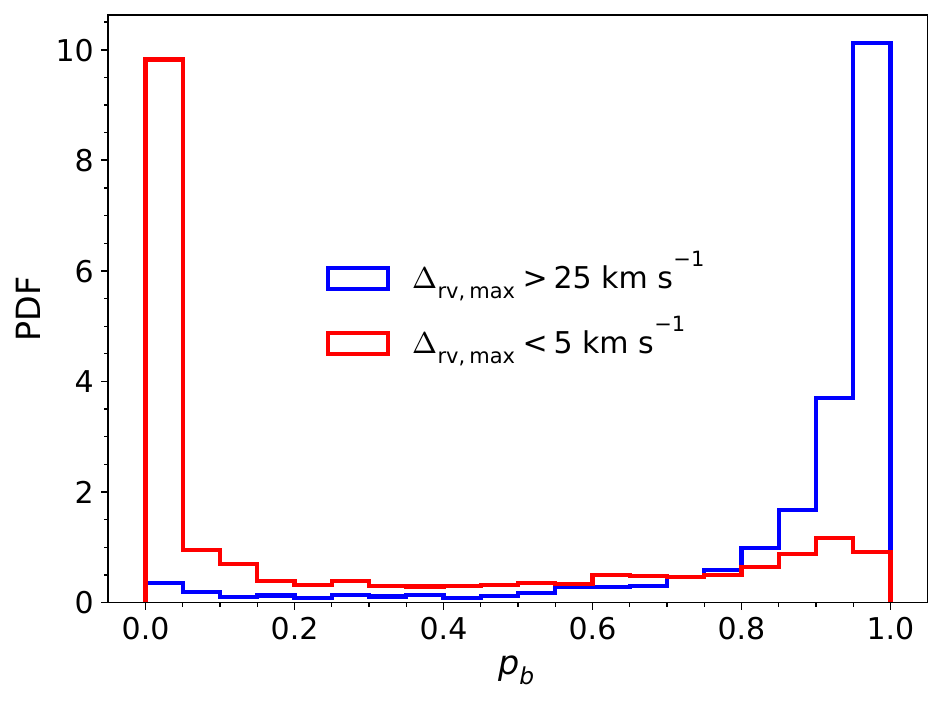}
	\caption{Probability density function (PDF) of the predicted binary probabilities ($p_b$) for stars observed multiple times in the LAMOST survey. The blue line indicates potential binary stars (with $\Delta_{\rm rv,max} > 25$ km s$^{-1}$), while the red dashed line indicates likely single stars (with $\Delta_{\rm rv,max} < 5$ km s$^{-1}$).
	}
	\label{fig:rv}
\end{figure}

In this subsection, we evaluate our network by comparing it with a sample of eclipsing binary stars and a sample based on radial velocity variations. 

Eclipsing binaries (EBs) are identified by the periodic variations in their apparent magnitude, known as the light curve, which are due to geometric attributes rather than spectral characteristics. Therefore, they serve as an excellent independent validation verification for our network. In this work, we used an sample of Algol (EA) type and W Ursa Majoris (EW) type eclipsing binary stars from the \citet{Chen2020}, which detected 350,000 eclipsing binaries from Zwicky Transient Facility. We cross-matched their sample with our main sequence stars sample, obtaining 535 EA and 2,724 EW common eclipsing binary stars. The blue and red solid lines in Figure~\ref{fig:EWL_EAL} depict the probability density function of $p_b$ for EA and EW binary stars, respectively. By setting $p_\mathrm{th}$ to 0.5, 96.45\% of EA and 96.40\% of EW eclipsing binary stars are correctly identified as binary stars, demonstrating the high performance of our network.

We subsequently compared our network's predictions with results derived from variations in radial velocity. In a binary star system, both stars revolve around a shared center of mass. Consequently, unless the orbital plane is perpendicular to our line of sight, the radial velocity observed from the system's spectrum will vary over time. In contrast, a single star's radial velocity will remain relatively constant. This distinction is utilized to identify binary stars \citep[e.g.][]{Price-Whelan2018, Birko2019, Qian2019}. The LAMOST survey has observed numerous stars on multiple occasions. For stars with at least two observations and a maximum radial velocity difference, $\Delta_{\rm rv,max} > 25 ~\rm{km~s}^{-1}$, we classify them as potential binary stars. Conversely, stars observed four or more times with $\Delta_{\rm rv,max} < 5~\rm{km~s}^{-1}$ are classified as likely single stars. We applied our network to both groups and plotted their probability density functions (PDFs) of $p_b$ in Figure~\ref{fig:rv}, where the black solid and cyan dashed lines represent the PDFs for possible binary and single stars, respectively. With a threshold probability $p_{\rm th}$ of 0.5, our network's prediction aligns with radial velocity variations for 92.8\% of binary stars and 68.8\% of single stars. We notice that the alignment for single stars is not as high as for binary stars, possibly because some stars in long-period binary systems exhibit minimal radial velocity changes, which LAMOST has not detected.

\subsection{Main sequence binary stars catalogue}

\begin{figure}
	\includegraphics[width=\columnwidth]{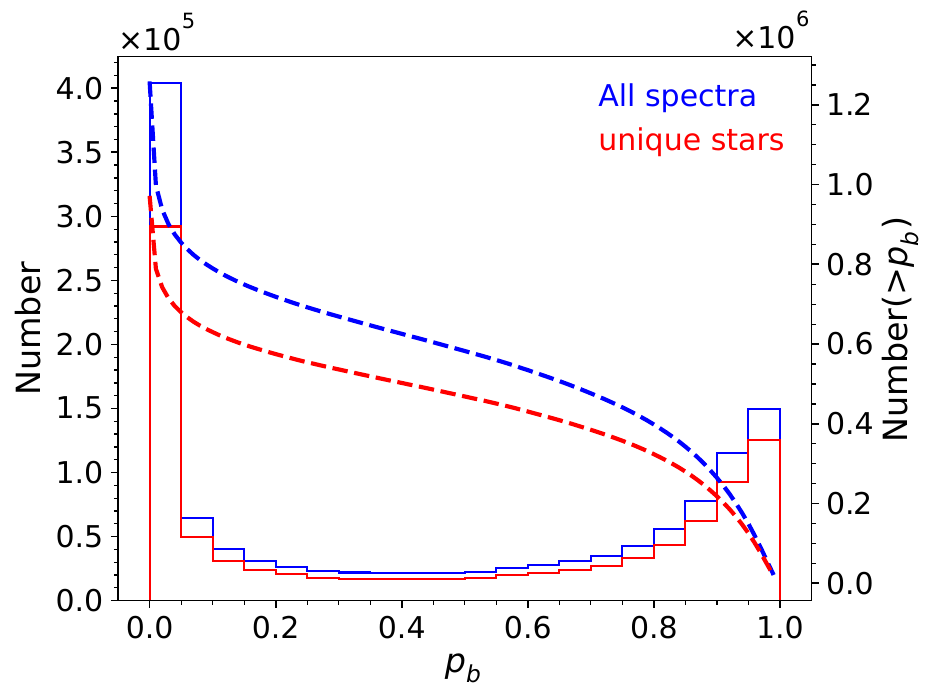}
	\caption{The distribution of $p_b$ provided by the network for all spectra (blue) and individual stars (red) within the main sequence star sample. The solid curves represent the number distribution of $p_b$, while the dashed lines indicate the count of spectra (or individual stars) with as a function of the $p_\mathrm{th}$. For individual stars, $p_b$ corresponds to the highest value in cases of multiple observations.}
	\label{fig:all_prob}
\end{figure}

\begin{figure}
	\includegraphics[width=\columnwidth]{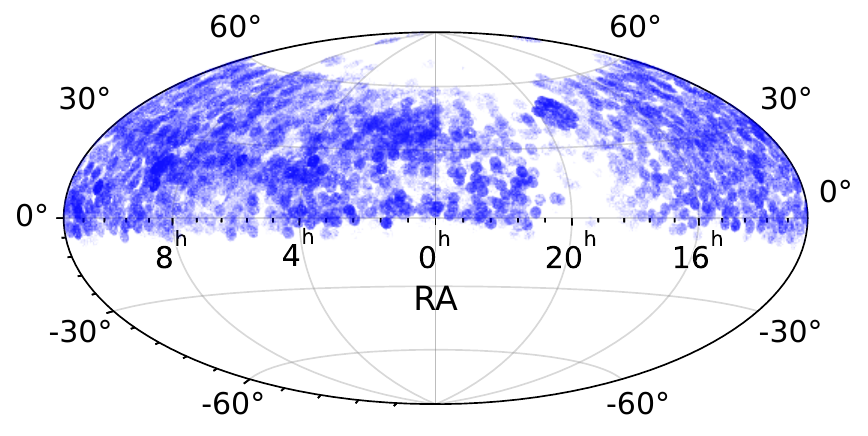}
	\caption{Distribution of binary stars in the Right Ascension (RA) and Declination (Dec) plane.}
	\label{fig:radec}
\end{figure}

\begin{figure}
	\includegraphics[width=\columnwidth]{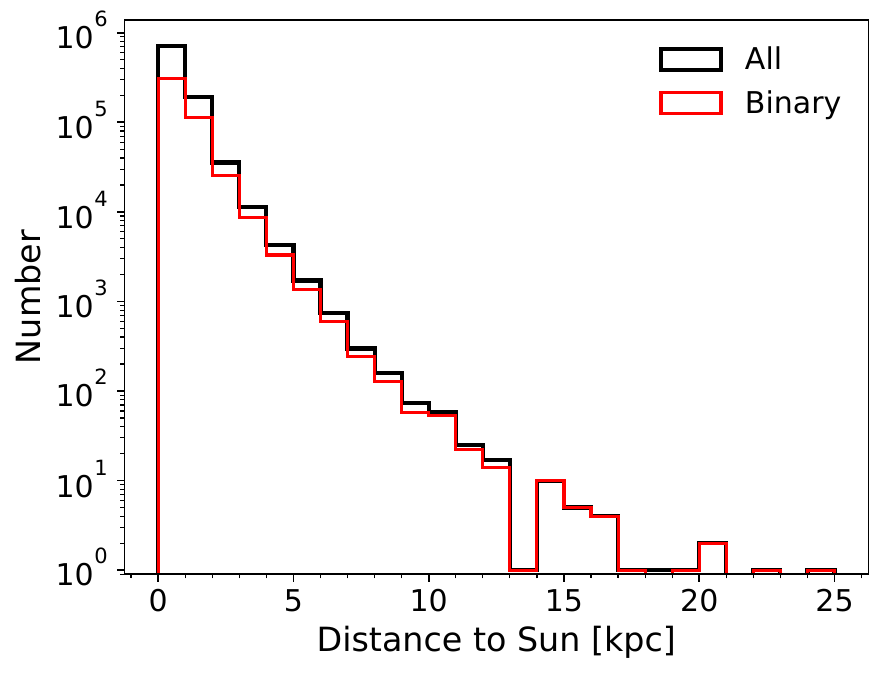}
	\caption{Distribution of distances to the Sun for all stars (black) and binary stars (red). The histogram reveals that binary stars are found at a range of distances, with 115 binary stars situated beyond 10 kpc.}
	\label{fig:dis}
\end{figure}

Following the evaluation of the network's performance, we applied it to our sample main sequence star sample. We selected spectra with $p_b > 0.5$ for incorporation into our primary catalog of binary stars on the main sequence (Table~\ref{tab:bs}). The catalog contains 468,634 stars\footnote{The training samples are included in this final binary stars catalog.}, with 323,909 of them having $p_b > 0.8$. The threshold selection is a compromise between accuracy and purity. Although we used a threshold of $p_\mathrm{th} = 0.5$ in this study, our published binary star catalog includes the $p_b$ value for each star to allow custom threshold selection based on specific requirements.

\citet{Qian2019} detected binary stars from LAMOST using a method based on variations in radial velocity from multiple observations of stellar spectra. Our network necessitates just a single observation of star spectra, but it is restricted to main sequence stars due to the limitations in the training dataset. They identified 256,000 spectroscopic binary or variable stars, slightly fewer than our findings. Our catalog surpasses prior spectroscopic binary star studies in size. For instance, \citet{Jack2019} identified 34,691 spectroscopic binary stars by comparing Gaia DR2 with other radial velocity catalogs, while \citet{Birko2019} discovered 27,716 SB1 from 450,646 stars with multiple radial velocity measurements in the RAVE and Gaia DR2 surveys. Additionally, \citet{Price-Whelan2018} found 4,898 SB1 amongst 96,231 red giant stars observed in the APOGEE survey.

We subsequently analyze the catalog's characteristics. Figure~\ref{fig:all_prob} illustrates the $p_b$ distribution, with blue and red solid lines representing all spectra and stars, respectively. For stars with multiple observational spectra, the $p_b$ value corresponds to the maximum value. The majority of the spectra (or unique stars) have $p_b$ values near zero or one, predicting as either single or binary star, without ambiguity. This shows a clear distinction in the binary status of the stars based on their $p_b$ values, making the classification straightforward. The dashed lines depict the number of spectra (or stars) as a function of threshold $p_\mathrm{th}$, indicating the count of those with $p_b$ exceeding $p_\mathrm{th}$. As $p_\mathrm{th}$ increases, the number decreases, yet the $Precision$ (see Figure~\ref{fig:test_prob}) of binaries improves.

\begin{table*}
\caption{Main Sequence Binary Star Candidates}
\label{tab:bs}
\centering
\begin{tabular}{cccccc}
\hline\hline
LAMOST-DR10-obsid & Gaia-DR3-source\_id & RA & DEC & $p_b$ & snrg\\
 & &($hms$)& ($dms$) & & \\
\hline
1411021 & 167617837532848128 & 03:59:05.55 & +30:37:10.4 & 0.971 & 30.180000 \\
1415160 & 167552897627512704 & 03:57:18.77 & +29:48:32.3 & 0.941 & 36.910000 \\
1415164 & 167551145280831232 & 03:58:10.41 & +29:49:59.4 & 0.981 & 25.450001 \\
1415222 & 167507130456235520 & 03:59:30.18 & +29:50:41.7 & 0.889 & 34.400002 \\
1415226 & 167106564626111744 & 03:58:33.72 & +29:22:49.9 & 0.912 & 28.770000 \\
... & ... & ... & ... & ... & ... \\
\hline
\multicolumn{5}{l}{\textit{Note:} Full catalog available in the online journal and via\dataset[DOI: 10.57760/sciencedb.15844]{https://doi.org/10.57760/sciencedb.15844}.} \\
\end{tabular}
\end{table*}

\begin{table*}
\caption{Main Sequence Binary Star Candidates Hosting Exoplanets}
\label{tab:exo_bs}
\centering
\begin{tabular}{cccccc}
\hline\hline
hostname &       LAMOST-DR10-obsid & Gaia-DR3-source\_id &           RA &          DEC &     $p_b$ \\
 & & &($hms$)& ($dms$) &\\
\hline
Kepler-971 & 243016079 & 2103712163117363200 & 18:56:22.73 & +41:10:58.0 & 0.895 \\
Kepler-743 & 247607014 & 2077994272606400512 & 19:42:17.80 & +42:48:23.2 & 0.756 \\
Kepler-478 & 247616169 & 2128256492468364160 & 19:29:56.86 & +46:11:46.4 & 0.704 \\
Kepler-1332 & 247710091 & 2126102411750601088 & 19:24:06.86 & +43:54:49.2 & 0.864 \\
Kepler-1583 & 249201170 & 2130695518492411904 & 19:05:32.86 & +47:01:00.0 & 0.960 \\
... &  ... &  ... &  ... &  ... &  ... \\
\hline
\multicolumn{6}{l}{\textit{Note:} Full catalog available in the online journal and via\dataset[DOI: 10.57760/sciencedb.15844]{https://doi.org/10.57760/sciencedb.15844}.} \\
\end{tabular}
\end{table*}

Figure~\ref{fig:radec} illustrates the distribution of binary spectra ($p_\mathrm{th}=0.5$) within the RA--DEC plane. This distribution closely resembles the overall spectra distribution (not depicted). In Figure~\ref{fig:dis}, we present the target distances from the Sun for total and binary stars, utilizing distances from \citet{Bailer2021}. Our analysis reveals that the range of distances stretches up to roughly 19 kpc, with a median distance of about 0.7 kpc for binary stars. There are 115 binary stars situated at distances greater than 10 kpc from the Sun.

\begin{figure}
	\includegraphics[width=\columnwidth]{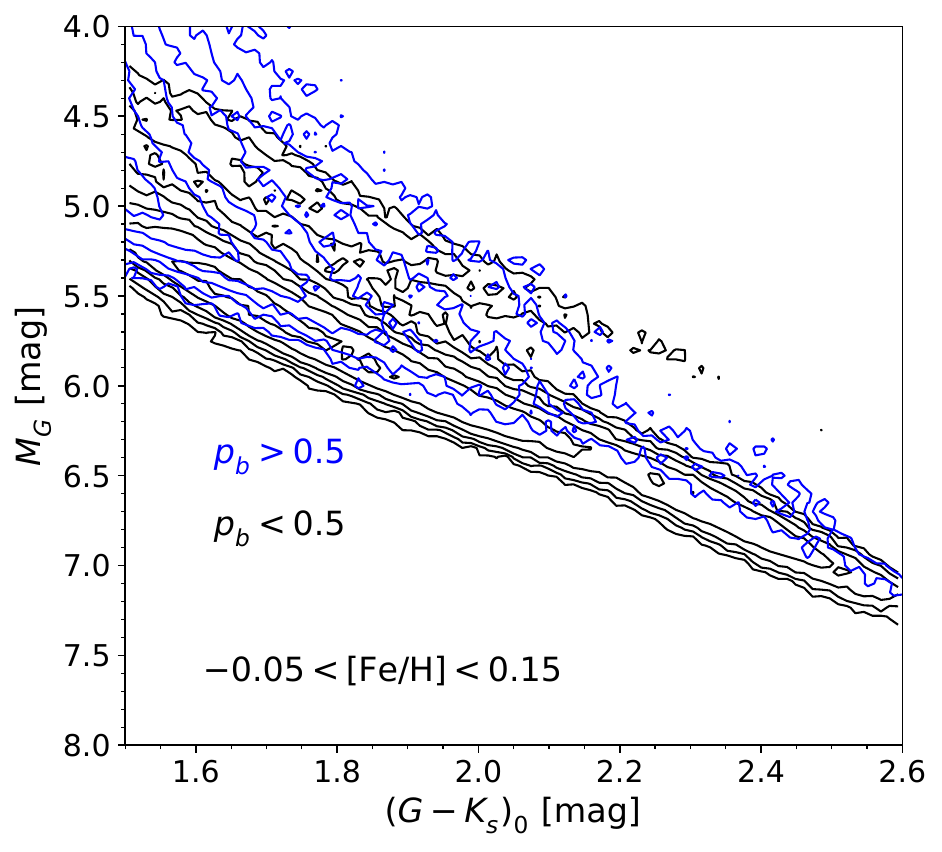}
	\caption{Number density contour plot in the color--magnitude diagram the stars with $-0.05< \mathrm{[Fe/H]} <0.15$. Stars with $p_b < 0.5$ are shown in black contours, while those with $p_b > 0.5$ are highlighted in blue.}
	\label{fig:HR}
\end{figure}

As the training sample differentiates single stars and binary stars in the color-magnitude diagram (CMD), it is valuable to examine the result from the network in the CMD. We plot single stars (chosen by $p_b<0.5$) and binary stars (chosen by $p_b>0.5$) within $-0.05< \mathrm{[Fe/H]} <0.15$ on the CMD in Figure~\ref{fig:HR}. The black and blue curves show the number density contour for single stars (chosen by $p_b<0.5$) and binary stars, respectively. Consistent with our expectations and similar to Figure~\ref{fig:HR_train}, the predicted binary stars appear above the single stars' main sequence in the CMD, even though they lack a clear boundary. We also observed a small fraction of stars with $p_b<0.5$ situated above binary stars. These stars are most likely binary stars with a high mass ratio (indicated by red points in Figure~\ref{fig:HR_train}). Since their spectra resemble that of single stars, they were excluded from the training dataset and consequently not identified by the network, which is as anticipated. Nevertheless, this does not affect the accuracy of our binary star catalog.

\subsection{Planets in binary stars system}

Previous studies have demonstrated the existence of planets orbiting binary star systems \citep[e.g.][]{Desidera2007, Mugrauer2009}, sparking significant interest in understanding planet formation and evolution within these dynamically complex environments \citep[e.g.][]{Thebault2015}. We cross-matched our catalog of binary stars with the NASA Exoplanet Archive
\citep{PSCompPars}\footnote{Accessed on 2024-09-24 at 02:31, returning 5759 rows.}\footnote{\url{https://exoplanetarchive.ipac.caltech.edu/cgi-bin/TblView/nph-tblView?app=ExoTbls&config=PSCompPars}}.
This yielded 128 binary star systems hosting confirmed exoplanets (Table~\ref{tab:exo_bs}). Remarkably, 114 of these systems were not previously identified as binary stars in the Exoplanet Archive, highlighting the contribution of our catalog. The catalog proves useful for future studies exploring the properties and formation of planets in binary star systems. These systems represent a diverse range of binary configurations and planetary architectures. Further analysis of this sample could shed light on the influence of stellar multiplicity on planet formation and evolution.

\section{Conclusions and discussions}
\label{sect:con}
In this study, we employed a convolutional neural network (CNN) to identify binary stars using low-resolution spectra from the LAMOST survey. Our CNN was trained on a dataset of 68,299 single stars and 3,818 intermediate mass-ratio binaries (between approximately 0.71 and 0.93) selected from \citet{Liu2019}, based on their distinct locations in the Hertzsprung-Russell diagram.

The performance of our network was rigorously evaluated using several methods. The area under the receiver operating characteristic curve (ROC) on a held-out test set reached 0.949, demonstrating a strong ability to differentiate between single and binary stars.  Furthermore, our network achieved a high detection rate of approximately 96\% when validated against a sample of known eclipsing binaries.  Comparison with radial velocity variations from multi-epoch LAMOST observations showed a 92.8\% agreement for binary stars and 68.8\% for single stars. The lower agreement for single stars might be attributed to undetected long-period binaries exhibiting minimal radial velocity changes within the LAMOST observation baseline.

Applying our trained CNN to the selected main sequence sample of 971,805 stars from LAMOST yielded a catalog of 468,634 binary stars, which are mainly intermediate mass ratio binaries given the training data.  The threshold of $p_b > 0.5$ for inclusion in the catalog represents a balance between maximizing the number of identified binaries and maintaining a high precision.  However, the full catalog includes the $p_b$ value for each star, allowing users to apply custom thresholds based on their specific requirements.  The catalog spans a distance range up to $\sim$19 kpc, with a median distance of $\sim$0.7 kpc for binary stars, and includes 115 binaries located beyond 10 kpc.  Interestingly, we also identified 128 binary systems hosting confirmed exoplanets by cross-matching our catalog with the NASA Exoplanet Archive.

A key advantage of our method is its ability to identify binary candidates using single-epoch spectra, unlike traditional radial velocity methods that require multiple observations. This makes our approach particularly valuable for large spectroscopic surveys like LAMOST.  However, it is important to acknowledge limitations.  Our training sample focused solely on main sequence stars, potentially limiting the applicability of the CNN to other stellar types.  Furthermore, potential biases in the training data may affect the network's performance. Future work will focus on expanding the training dataset to include a wider range of stellar types and exploring alternative methods for training data selection to mitigate potential biases. 

This large catalog of binary star candidates provides a valuable resource for future research on the formation, evolution, and statistical properties of binary systems, particularly for studies requiring large sample sizes. The catalog, along with individual values $p_b$, is available at https://doi.org/10.57760/sciencedb.15844.


\section*{Acknowledgments}
We would like to thank the referee for the constructive
suggestions and comments. This work was supported by the China National Key Program for Science and Technology Research and Development of China (2022YFA1602901, 2023YFA1608204), the National Natural Science Foundation of China (Nos. 11988101, 11873051, 12125302, 12041305, 12173016, 12203065), the CAS Project for Young Scientists in Basic Research grant (No. YSBR-062), and the K.C. Wong Education Foundation, and the science research grants from the China Manned Space Project. We acknowledge the CSST research grants from the China Manned Space Project.
Y.J. acknowledges support from the Cultivation Project for FAST Scientific Payoff and Research Achievement of CAMS-CAS.
JW acknowledges the hospitality of the International Centre of Supernovae (ICESUN), Yunnan Key Laboratory at Yunnan Observatories Chinese Academy of Sciences when we drafted this paper.

Guoshoujing Telescope (the Large Sky Area Multi-Object Fiber Spectroscopic Telescope LAMOST) is a National Major Scientific Project built by the Chinese Academy of Sciences. Funding for the project has been provided by the National Development and Reform Commission. LAMOST is operated and managed by the National Astronomical Observatories, Chinese Academy of Sciences.

This work has made use of data from the European Space Agency (ESA) mission
{\it Gaia} (\url{https://www.cosmos.esa.int/gaia}), processed by the {\it Gaia}
Data Processing and Analysis Consortium (DPAC,
\url{https://www.cosmos.esa.int/web/gaia/dpac/consortium}). Funding for the DPAC
has been provided by national institutions, in particular the institutions
participating in the {\it Gaia} Multilateral Agreement.

This publication makes use of data products from the Two Micron All Sky Survey, which is a joint project of the University of Massachusetts and the Infrared Processing and Analysis Center/California Institute of Technology, funded by the National Aeronautics and Space Administration and the National Science Foundation.

\bibliography{main}{}
\bibliographystyle{aasjournal}



\end{document}